\documentclass[10pt]{iopart}
\usepackage{iopams}
\usepackage{graphicx}
\usepackage{nicefrac}

\newcommand{\Ref}[1]{(\ref{#1})}

\newcommand{\mean}[1]{\langle #1 \rangle}
\def\bbL{{\mathbb L}}

\usepackage{comment}

\begin{document}
\title{On the universality of knot probability ratios}
\author{E.J.~Janse~van~Rensburg}
\address{Department of Mathematics and Statistics, York University}
\ead{rensburg@yorku.ca}
\author{A.~Rechnitzer}
\address{Department of Mathematics, University of British Columbia}
\ead{andrewr@math.ubc.ca}

\date{\today}

\begin{abstract}
Let $p_n$ denote the number of self-avoiding polygons of length $n$ on a regular three-dimensional
lattice, and let $p_n(K)$ be the number which have knot type $K$. The
probability that a random polygon of length $n$ has knot type $K$ is
$p_n(K)/p_n$ and is known to decay exponentially with length
\cite{Sumners1988,Pippenger1989}. Little is known rigorously about
the asymptotics of $p_n(K)$, but there is substantial numerical evidence 
\cite{Orlandini1998, Marcone2007,Rawdon2008, JvR2008} that $p_n(K)$ grows as
\vspace{2ex}
\begin{eqnarray*}
  p_n(K) &\simeq& \, C_K \, \mu_\emptyset^n \, n^{\alpha-3+N_K}, \quad \mbox{as
$n \to \infty$},
\end{eqnarray*}\\[-1ex] \noindent
where $N_K$ is the number of prime components of the knot type $K$. It is
believed that the 
entropic exponent, $\alpha$, is universal, while the exponential growth rate,
$\mu_\emptyset$, is independent of the knot type but varies with the lattice.
The amplitude, $C_K$, depends on both the lattice and the knot type.

The above asymptotic form implies that the relative probability of a random
polygon of length $n$ having prime knot type $K$ over prime knot type $L$ is
\vspace{1ex}
\begin{eqnarray*}
  \frac{p_n(K)/p_n}{p_n(L)/p_n} &=& \frac{p_n(K)}{p_n(L)} 
   \simeq \left[ \frac{C_K}{C_L} \right].
\end{eqnarray*}\\[-1ex] \noindent
In the thermodynamic limit this probability ratio becomes an amplitude ratio; it
should  be universal and depend only on the knot types $K$ and $L$. In this
letter we examine the universality of these probability ratios for polygons in
the simple cubic, face-centered  cubic, and body-centered cubic lattices.  Our
results support the hypothesis that these are universal quantities. 
For example, we estimate that a long random polygon is approximately 28 times
more likely to be a trefoil than be a figure-eight, independent of the
underlying lattice, giving an estimate of the intrinsic entropy associated
with knot types in closed curves. 

\end{abstract}



\submitto{\JPA}

\maketitle

\section{Introduction}
A self-avoiding polygon (SAP) on a regular lattice $\bbL$ is the piecewise linear embedding of 
a simple closed curve as a sequence of distinct edges joining vertices in $\bbL$.  The number of
distinct unrooted polygons modulo translations is denoted $p_n$. It is known that $\lim_{n\to\infty} 
p_n^{\nicefrac{1}{n}} = \mu$ exists, where $\mu$ is the self-avoiding walk
growth constant 
\cite{Hammersley1960, Hammersley1961}.

\begin{figure}[h!]
 \begin{center}
  \includegraphics[height=3.6cm]{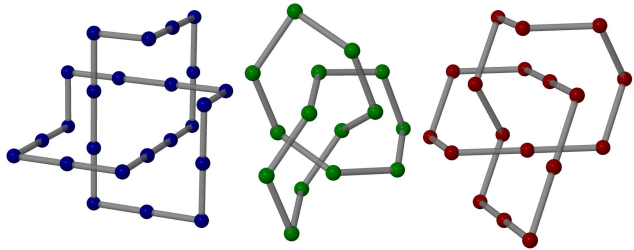}
 \end{center}
\vspace{-2ex}
  \caption{Minimal trefoils on the SC (left --- 24 edges), FCC (middle --- 15
edges) and BCC (right --- 18 edges).}
  \label{fig sc31}
\end{figure}

The knot type of polygons in three-dimensional lattices are well defined.  Denote by $p_n(K)$
the number of unrooted polygons of length $n$ and knot type $K$, modulo translations. Computing
$p_n$ or $p_n (K)$ is a very difficult combinatorial problem, though determining the minimal
length $n$ such that $p_n(K)>0$ and the numbers of shortest embeddings is viable for knot types of
low complexity.  For example, there are 3 shortest unknots of length 4 in the simple cubic 
lattice (SC), 8 of length 3 in the face-centred cubic lattice (FCC), and 12 of
length 4 in the body-centered cubic lattice (BCC). The simplest non-trivial knot
type is the trefoil (denoted by $3_1$, see Figure~\ref{fig sc31}) and it is
known that $p_n (3_1) = 0$ if $n<24$ and $p_{24} (3_1) = 3328$ in the SC
\cite{Scharein2009}. Data on polygons collected by the GAS algorithm 
\cite{JvR2010a} shows that $p_{15} (3_1) = 128$ in the FCC and $p_{18} (3_1) = 3168$ in
the BCC (see Table~\ref{tab min knot}); no shorter trefoils were observed.

Numerical studies \cite{Orlandini1998, Marcone2007, JvR2008,Rawdon2008, JvR2010a} have shown 
that $p_n(K)$ behaves as
\vspace{1ex}\begin{equation}
  p_n(K) \simeq \, C_K \, \mu_\emptyset^n \, n^{\alpha-3+N_K}, \quad \mbox{as
$n \to
\infty$},
\label{eqn pnk}
\end{equation}
where  $N_K$ is the number of prime components of the knot type $K$. The exponent is
thought to be universal, while the growth rate, $\mu_\emptyset$, depends on the
lattice but not the knot type. The amplitude, $C_K$, depends on both the lattice
and the knot type. Unfortunately very little of this form can be proven
rigorously --- the exponential growth rate is only known to exist when $K$ is
the unknot. A pattern-theorem \cite{Sumners1988,Pippenger1989} shows that the
growth rate of unknots, $\mu_\emptyset$, is strictly smaller than $\mu$. The
same argument also shows that the probability that a polygon of length $n$ has
knot type $K$, given by $p_n(K)/p_n$, decays exponentially with length. 

In this paper we consider the asymptotic behaviour of ratios of knotting probabilities. In
particular, for two prime knot types $K$ and $L$ one has $N_K = N_L=1$ and the ratio 
of probabilities is given by
\begin{equation}
\frac{p_n(K)/p_n}{p_n(L)/p_n} = \frac{p_n(K)}{p_n(L)} \simeq \left[ \frac{C_K}{C_L} \right],
\quad\hbox{as $n\to\infty$}.
\end{equation}
Hence, the limiting ratio of probabilities approaches a constant. Since this limit is an 
\emph{amplitude ratio} we expect it to be universal --- depending only
on the knot types and the universality class of the underlying model.

Such ratios were studied previously on the SC \cite{JvR2010a} (by the methods
used in this paper) and in \cite{Baiesi2010} (by very different methods). Here,
we use the GAS algorithm to estimate $p_n(K)$ for various prime knots on the
SC, FCC and BCC lattices. Our results indicate that the above ratio is dependent
on the knot types, but independent of the underlying lattices, and
so, universal.

\section{Atmospheric moves on cubic lattices}
The GAS algorithm \cite{JvR2009a} samples along sequences of conformations that evolve
through local elementary transitions called atmospheric moves (see eg. \cite{JvR2009b}). 
The algorithm is a generalisation of the Rosenbluth algorithm
\cite{Hammersley1954, Rosenbluth1955}, and is an approximate enumeration
algorithm.

The GAS algorithm was used to estimate the number of knotted polygons on the SC
\cite{JvR2010a} using BFACF moves \cite{Berg1981,AdCC1983,AdCCF1983} as
atmospheric moves. This implementation relies on a result in \cite{JvR1991}
that the irreducibility classes of the BFACF  elementary moves applied to SC
polygons are the classes of polygons of fixed knot types~\cite{JvR1991}.
%
\begin{figure}[h!]
 \begin{center}
  \includegraphics[height=1.4cm]{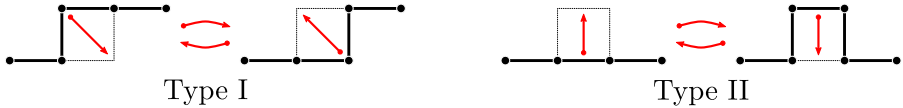}
 \end{center}
\vspace{-2ex}
\caption{Elementary moves of the BFACF algorithm on the SC lattice.}
\label{fig bfacf moves}
\end{figure}

BFACF elementary moves (see Figure~\ref{fig bfacf moves}) are either neutral (or Type~I) operating 
on two adjacent orthogonal edges of a SC polygon, or the are of Type~2 which
are positive or negative length changing moves. A neutral moves exchanges two
adjacent edges over a unit lattice 
square which defines a \textit{neutral atmospheric plaquette}. A \textit{negative move} replaces 
three edges in a $\sqcap$ conformation by a single edge and so defines a \textit{negative 
atmospheric plaquette}. Similarly a \textit{positive move} replaces a single edge of the polygon 
by three edges in a $\sqcap$ arrangement; these edges define a 
\textit{positive atmospheric plaquette}. Let $a_+(\varphi), a_0(\varphi),a_-(\varphi)$ be  
the total numbers positive, neutral and negative atmospheric moves of a SC lattice polygon $\varphi$.

\begin{figure}[h!]
 \begin{center}
  \includegraphics[height=3.3cm]{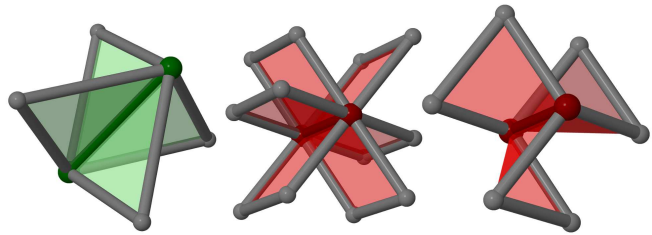}
 \end{center}
  \vspace{-2ex}
\caption{(Left) There are 4 elementary triangular plaquettes incident to each
edge in the FCC lattice. (Middle and right) Each edge in the BCC lattice is
incident to 12 plaquettes; 6 planar and 6 non-planar (the remaining 3 are
reflections of the 3 displayed).}
\label{fig plaq}
\end{figure}

The plaquettes in the FCC and BCC lattice (see Figure~\ref{fig plaq}) define elementary moves 
in the FCC and BCC analogous to the BFACF moves. Since the FCC plaquettes are triangles they 
define positive and negative moves, while the BCC plaquettes are quadrilaterals and so also 
give neutral moves. These generalisations are discussed at length in \cite{JvR2010b} and it  
is shown that on each lattice the irreducibility classes of the moves coincide with classes 
of polygons of fixed knot types.

\section{GAS sampling of knotted polygons}
We have implemented the GAS algorithm using the atmospheric moves described
above. Let $\varphi_0$ be a lattice polygon, then the GAS algorithm samples
along a sequence of polygons $(\varphi_0, \varphi_1, \dots)$, where
$\varphi_{j+1}$ is obtained from $\varphi_j$ by an atmospheric move.

Each atmospheric move is chosen uniformly from the  possible moves, so that if $\varphi_j$ has 
length $\ell_j$ then
\begin{equation}
  \Pr(\mbox{$+$}) 
  \propto \beta_{\ell_j} 
  a_+(\varphi_j),\quad
  \Pr(\mbox{$0$})
  \propto 
  a_0(\varphi_j),\quad
  \Pr(\mbox{$-$})
  \propto 
  a_-(\varphi_j) .
\end{equation}
where $\beta_\ell$ is a parameter that is chosen to  be approximately 
$\frac{\mean{a_+}_{\ell} }{ \mean{a_-}_{\ell} }$.  This parameter can be chosen so that on 
average the probability of making a positive move is roughly  the same as that of making a 
negative move. This produces a sequence $\langle\varphi_j \rangle$ of states and
we assign a weight to each state:
\begin{eqnarray}
  W(\varphi_n) &=
  \frac{a_-(\varphi_0) + a_0(\varphi_0) + \beta_{\ell_0} a_+(\varphi_0) }
  {a_-(\varphi_{n}) + a_0(\varphi_{n}) + \beta_{\ell_{n}} a_+(\varphi_{n}) }
  \times
  \prod_{j=0}^n \beta_{\ell_j}^{(\ell_j - \ell_{j+1})}.
\end{eqnarray}

The probabilities and weights are functions of the number of possible atmospheric moves and so 
the algorithm must recalculate these efficiently. Since the elementary moves only involve 
local changes, executing a move and updating the polygon takes $O(1)$ time.

The resulting data were analysed by computing the mean weight $\mean{W}_n$ of polygon of length 
$n$ edges and then using the result (from \cite{JvR2009a})
\begin{eqnarray}
  \frac{\mean{W}_n}{\mean{W}_m} &= \frac{p_n(K)}{p_m(K)}.
  \label{eqn w ratio}
\end{eqnarray}
This gives approximations to the number of polygons of any given length $n$,
provided the number of polygons is known exactly at another length $m$.

\section{Results}
We collected data on the prime knots $3_1,4_1,5_1$ and $5_2$ on the three lattices. In order to 
use equation~\Ref{eqn w ratio} we computed the total number minimal length polygons of fixed 
given knot type --- see Table~\ref{tab min knot}. We did this by collecting them while performing 
the simulation (or in independent runs); this idea was used in~\cite{Scharein2009} 
and~\cite{JvR2010a} and our SC results agree. Typically, the algorithm quickly
found all realisations of minimal knots (within hours) and then failed to find
new conformations after  another few days of CPU time. We note that the result
for trefoils in the SC has been proved 
\cite{Diao1994, Scharein2009}.

\begin{table}[h!]
\begin{center}
 \begin{tabular}{||c||c|c||c|c||c|c||}
 \hline
  Knot & \multicolumn{2}{|c||}{SC} &
  \multicolumn{2}{|c||}{FCC} &
  \multicolumn{2}{|c||}{BCC} \\
  \hline
   & length & number 
   & length & number 
   & length & number 
   \\
   \hline
  $0_1$
  & 4 & 3 
  & 3 & 8 
  & 4 & 12 
  \\
  \hline \hline
  $3_1$
  & 24 & 3328 
  & 15 & 64 
  & 18 & 1584 
  \\
  \hline
  $4_1$
  & 30 & 3648 
  & 20 & 2796 
  & 20 & 12 
  \\
  \hline
  $5_1$
  & 34 & 6672 
  & 22 & 96 
  & 26 & 14832 
  \\
  $5_2$
  & 36 & 114912 
  & 23 & 768 
  & 26 & 4872 
  \\
  \hline
 \end{tabular}
\end{center}
 \caption{The number of minimal length polygons of fixed knot types in the SC,
FCC and BCC lattices.}
  \label{tab min knot}
\end{table}

Using the data in Table~\ref{tab min knot} and equation~\Ref{eqn w ratio} we were able to 
estimate $p_n(K)$ for each knot type in each of the three lattices. Each simulation ran for 
1 week on a single node of WestGrid's Glacier 
cluster\footnote{See \texttt{http://www.westgrid.ca}}. The implementation was particularly
simple and efficient in the FCC lattice and the SC simulations were faster than
the BCC lattice simulations.  Each simulation was composed of approximately
$400$ chains of length $2^{27}$ polygons on the FCC lattice, $1400$ chains of
$2^{23}$ polygons on the simple cubic  lattice and $500$ chains of $2^{23}$
polygons on the BCC lattice.  In each simulation we limited the maximum length
of the polygons to $512$ edges.

\begin{figure}[h!]
\begin{center}
\includegraphics[height=65mm]{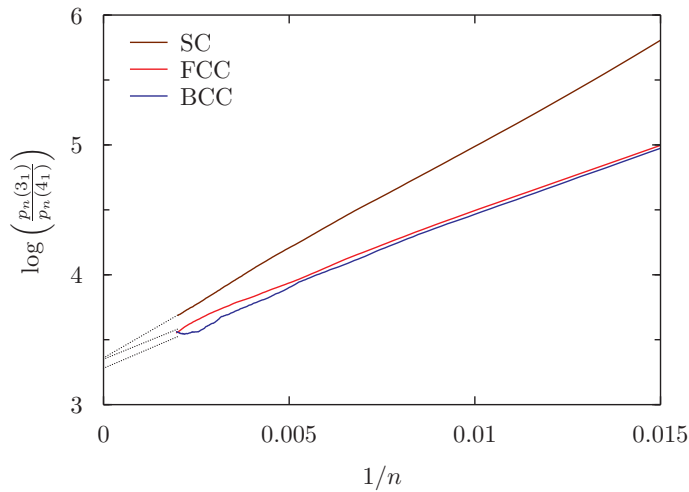}
\end{center}
\vspace{-5mm}
\caption{Plots of the logarithm of the ratio of the number of $3_1$ to $4_1$ knots. 
The dotted lines indicate the extrapolations. Note that the FCC and BCC data are 
nearly the same. The intercept indicates that the limiting ratio is 
approximately $e^{3.32}\approx 28$.}
\label{fig ratio 3141}
\end{figure}

\begin{figure}[h!]
\begin{center}
\includegraphics[height=65mm]{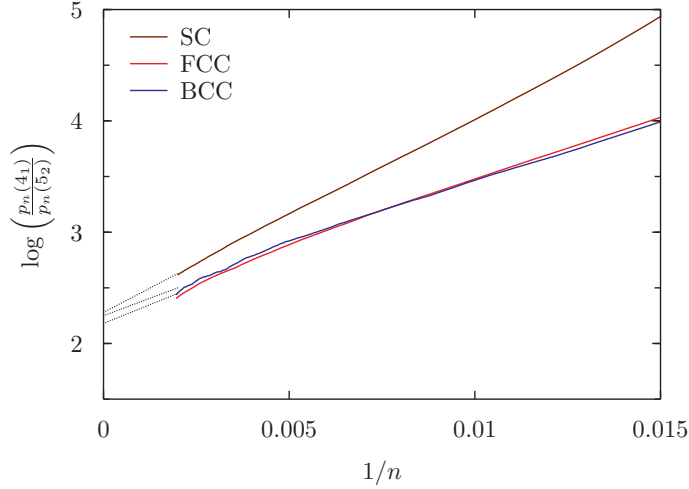}
\end{center}
\vspace{-5mm}
\caption{Plots of the logarithm of the ratio of the number of $4_1$ to $5_2$ knots. 
The linear extrapolations are indicated by dotted lines. The intercept indicates 
that the limiting ratio is approximately $e^{2.2} \approx 9$.}
\label{fig ratio 4152}
\end{figure}

The estimates of $p_n (K)$ in each lattice were used to extrapolate the ratios $p_n(K) / p_n(L)$ 
for fixed prime knots $K,L$. In earlier work on SC polygons \cite{JvR2010a} we
observed that the logarithm of these ratios were approximately linear in
$n^{-1}$. In Figures~\ref{fig ratio 3141}, \ref{fig ratio 4152} and~\ref{fig
ratio 5152} we plot the logarithm of the ratio against $n^{-1}$ for various
pairs of prime knots.

In Figure~\ref{fig ratio 3141} we show that there is strong agreement between the FCC and 
BCC data.   In addition, the three extrapolated curves appear to have approximately the 
same limit.  This is strong numerical support for the hypothesis that the limiting ratio 
is universal. 

Linear fits of the data gives $y$-intercepts of $3.34(3)$, $3.35(3)$, $3.29(3)$ 
for the SC, the FCC and the BCC lattices (respectively). These results includes
each other 
within $95$\% confidence intervals.  Exponentiating these results estimates the limiting 
amplitude ratio to be $27\pm 2$. If we exclude the BCC data, since it is not as well
converged at large $n$, we obtain a limiting ratio of $28 \pm 1$. 

Turning to the ratio of $4_1$ to $5_2$ plotted in Figure~\ref{fig ratio 4152} we find
similar results, though the data are not as well converged and our estimates are 
not as good. Linear fits lead to an estimate of $9\pm 1$ for the limiting ratio. 
The estimates on all three lattices agree, supporting the hypothesis of universality.
In the final plot we show the ratio of $5_1$ to $5_2$ knots. Again we find similar
results and estimate the limiting ratio to be $0.67(3)$. 

\begin{figure}[h!]
\begin{center}
\includegraphics[height=65mm]{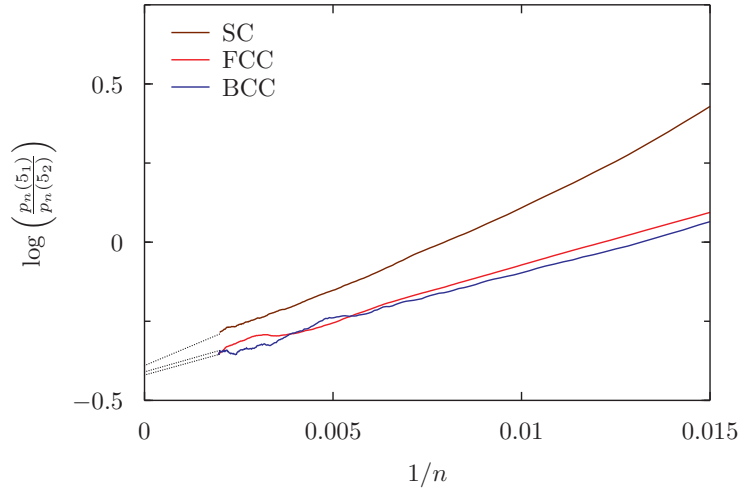}
\end{center}
\vspace{-5mm}
\caption{Plots of the logarithm of the ratio of the number of $5_1$ to $5_2$
knots. 
The linear extrapolations are indicated by dotted lines. The intercept indicates 
that the limiting ratio is approximately $e^{-0.4} \approx 0.67$.}
\label{fig ratio 5152}
\end{figure}

We have also studied the other ratios and find similar support for their universality. In summary:
\begin{eqnarray}
  \begin{array}{rlrlrl}
   ×
  \nicefrac{C_{3_1}}{C_{4_1}}&= 28(1)
  &\nicefrac{C_{3_1}}{C_{5_1}} &= 400(20)
  & \nicefrac{C_{3_1}}{C_{5_2}} &= 280(20)  \\[1ex]
  && \nicefrac{C_{4_1}}{C_{5_1}} &= 15(1)
  & \nicefrac{C_{4_1}}{C_{5_2}} &=9(1)  \\[1ex]
  &&&& \nicefrac{C_{5_1}}{C_{5_2}} &= 0.67(3).
  \end{array}
\end{eqnarray}
These numbers are self-consistent within the stated error bars.
Curiously, in each case we found that the curves for the FCC and BCC lie
close together while that of the SC stands apart; we would like to understand
this better, but have no explanation at this time.

  Some caution is needed when comparing these results to previous studies of
knot probability amplitudes such as \cite{Deguchi1997, Millett2005, Baiesi2010}.
Any estimate of the amplitude will have sensitive dependence on the estimate of
the exponent. Indeed, unless the estimated exponents are equal, the ratio of the
estimated probabilities will tend to zero or infinity.

Mindful of this, we may compare the ratio of estimated amplitudes for SC
polygons from \cite{Baiesi2010} we find $C_{3_1} / C_{4_1} \approx 22$ which is
close to our estimate, but not within mutual error bars. The ratio of estimated
amplitudes for off-lattice polygons from \cite{Deguchi1997}
and~\cite{Millett2005} give quite different results.  However, it is not clear
that our models are in the same universality class as these off-lattice models. 
In addition, the comparison may also be affected by differences in estimated
entropic exponents in these studies.

\section{Conclusions}
We have studied the ratio of probabilities of different knot types. The
scaling assumption in equation~\Ref{eqn pnk} indicates that the limit of this
ratio should be an amplitude ratio and thus universal.  Using the GAS algorithm
we have formed direct estimates of the number of polygons of various prime knot
types on three different lattices.  Extrapolating from these estimates provides
numerical evidence that the probability ratios are universal --- depending only
on the knot types and the universality class of the underlying model. In
particular we find that a long polygon is about 28 times more likely to be a
trefoil than a figure-eight.

There are a number of extensions of this work that we would like to pursue ---
extending these results to composite knots and links, and also to perform
similar analyses of data from off-lattice models.

\section*{Acknowledgements}
We would like to thank BIRS and the organisers of the conference where we had the idea
for this paper after discussions with several participants; in particular, Bertrand
Duplantier. We are also indebted to Stu Whittington, Thomas Prellberg and Enzo
Orlandini for their careful reading of the manuscript.  Additionally we would
like to thank the anonymous referees for their helpful suggestions. The
simulations were run on the WestGrid computer cluster and we thank them for
their support. Finally, both authors acknowledge financial support from NSERC,
Canada.

\section*{References}
\bibliographystyle{unsrt}
\bibliography{biblio}

\end{document}